\documentclass[preprint2]{aastex631}
\begin{document}

\title{An Overview of Solar Orbiter Observations of Interplanetary Shocks in Solar Cycle 25}

\newcommand{\tbn}{$\theta_{Bn}$}

\correspondingauthor{Domenico Trotta}
\email{d.trotta@imperial.ac.uk}

\author[0000-0002-0608-8897]{Domenico Trotta}
\affiliation{European Space Agency (ESA), European Space Astronomy Centre (ESAC), Camino Bajo del Castillo s/n, 28692 Villanueva de la Cañada, Madrid, Spain}
\affiliation{The Blackett Laboratory, Department of Physics, Imperial College London, London SW7 2AZ, UK}

\author[0000-0003-1589-6711]{Andrew Dimmock}
\affiliation{Swedish Institute of Space Physics, 751 21 Uppsala, Sweden}

\author[0000-0002-3039-1255]{Heli Hietala}
\affiliation{Department of Physics and Astronomy, Queen Mary University of London, London E1 4NS, UK}

\author[0000-0001-7171-0673]{Xochitl Blanco-Cano}
\affiliation{Departamento de Ciencias Espaciales, Instituto de Geofísica, Universidad Nacional Autónoma de México, Ciudad Universitaria, 04150 Ciudad de México, Mexico}

\author[0000-0002-7572-4690]{Timothy S. Horbury}
\affiliation{The Blackett Laboratory, Department of Physics, Imperial College London, London SW7 2AZ, UK}

\author[0000-0002-0074-4048]{Rami Vainio}
\affiliation{Department of Physics and Astronomy, University of Turku, FI-20014 Turku, Finland}

\author[0000-0003-3903-4649]{Nina Dresing}
\affiliation{Department of Physics and Astronomy, University of Turku, FI-20014 Turku, Finland}

\author[0000-0002-0606-7172]{Immanuel Christopher Jebaraj}
\affiliation{Department of Physics and Astronomy, University of Turku, FI-20014 Turku, Finland}

\author[0000-0001-9039-8822]{Francisco Espinosa}
\affiliation{Universidad de Alcalá, Space Research Group,
28805 Alcalá de Henares, Spain}

\author[0000-0002-5705-9236]{Ra\'ul G\'omez-Herrero}
\affiliation{Universidad de Alcalá, Space Research Group,
28805 Alcalá de Henares, Spain}

\author[0000-0002-4240-1115]{Javier Rodriguez-Pacheco}
\affiliation{Universidad de Alcalá, Space Research Group,
28805 Alcalá de Henares, Spain}

\author[0000-0002-0623-6992]{Yulia Kartavykh}
\affiliation{Institute of Experimental and Applied Physics, Kiel University, D-24118 Kiel, Germany }

\author[0000-0002-3176-8704]{David Lario}
\affiliation{Heliophysics Science Division, NASA Goddard Space Flight Center, Greenbelt, MD 20771, USA}

\author[0000-0003-1848-7067]{Jan Gieseler}
\affiliation{Department of Physics and Astronomy, University of Turku, FI-20014 Turku, Finland}

\author[0000-0002-6203-5239]{Miho Janvier}
\affiliation{European Space Agency, ESTEC, Noordwijk, The Netherlands}

\author{Milan Maksimovic}
\affiliation{LESIA, Observatoire de Paris, Université PSL, CNRS, Sorbonne Université, Univ. Paris Diderot, Sorbonne Paris Cité, 5 Place Jules Janssen, 92195 Meudon, France}

\author[0000-0002-9774-9047]{Nasrin Talebpour Sheshvan}
\affiliation{Department of Physics and Astronomy, University of Turku, FI-20014 Turku, Finland}

\author{Christopher J. Owen}
\affiliation{Department of Space and Climate Physics, Mullard Space Science Laboratory, University College London, Dorking, Surrey, RH5 6NT, UK}

\author[0000-0002-4489-8073]{Emilia K. J. Kilpua}
\affiliation{Department of Physics, University of Helsinki, FI-00014 Helsinki, Finland}

\author[0000-0002-7388-173X]{Robert F. Wimmer-Schweingruber}
\affiliation{Institute of Experimental and Applied Physics, Kiel University, D-24118 Kiel, Germany}



\begin{abstract}
Interplanetary shocks are fundamental constituents of the heliosphere, where they form as a result of solar activity. We use previously unavailable measurements of interplanetary shocks in the inner heliosphere provided by Solar Orbiter, and present a survey of the first 100 shocks observed in situ at different heliocentric distances during the rising phase of solar cycle 25. The fundamental shock parameters (shock normals, shock normal angles, shock speeds, compression ratios, Mach numbers) have been estimated and studied as a function of heliocentric distance, revealing a rich scenario of configurations. Comparison with large surveys of shocks at 1~au show that shocks in the quasi-parallel regime and with high speed are more commonly observed in the inner heliosphere. The wave environment of the shocks has also been addressed, with about 50\% of the events exhibiting clear shock-induced upstream fluctuations. We characterize energetic particle responses to the passage of IP shocks at different energies, often revealing  complex features arising from the interaction between IP shocks and pre-existing fluctuations, including solar wind structures being processed upon shock crossing. Finally, we give details and guidance on the access use of the present survey, available on the EU-project ``solar energetic particle analysis platform for the inner heliosphere'' (SERPENTINE) website. The algorithm used to identify shocks in large datasets, now publicly available, is also described. 
\end{abstract}



\section{Introduction} \label{sec:intro}

Shocks are ubiquitously observed in astrophysical environments, where they are believed to play a crucial role in energy conversion and particle acceleration~\citep[e.g.][]{Bykov2019}. Despite decades of research, the mechanisms by which shocks mediate such processes of energy conversion and particle acceleration are still a matter of debate~\citep{Lee2012}. In general, shocks are abrupt transitions between supersonic and subsonic flows, converting directed bulk flow energy (upstream) into heat and magnetic energy (downstream)~\citep{Marcowith2016}. In the collisionless case, a fraction of the available energy can be channelled in the production of energetic particles~\cite [e.g.,][]{Drury1983}.      

Heliospheric shocks are unique as accessible by direct spacecraft observations, and thus represent the missing link to astrophysical systems only observable remotely, like in the case of spectacular radiation emission due to shock--accelerated particles in supernova remnants~\citep[e.g.][]{Giuffrida2022}.  Most of our knowledge is built around direct observations of the Earth's bow shock, resulting from the interaction between the supersonic solar wind and the Earth's magnetosphere, which represents an obstacle to its propagation~\citep{Eastwood2015}. Since the early predictions and evidence due to the IMP8 mission~\citep{Dungey1979} to the modern NASA Magnetospheric MultiScale mission~\citep[MMS;][]{Burch2016} elucidating the details of how energy is partitioned across the shock transition~\citep{Schwartz2022}, the Earth's bow shock has been an invaluable resource to understand shock behaviour down to the smallest, kinetic scales. In the past decades,  particles reflected by the Earth's shock and the fluctuation they induce in the upstream plasma (namely particle and wave foreshocks) have been extensively documented using the large spacecraft fleet now orbiting Earth~\citep[e.g.,][]{Wilkinson2003,Wilson2016}, often combined with numerical efforts~\citep[e.g.,][]{Kartavykh2013, Turc2023}.

Interplanetary (IP) shocks travel in the heliosphere driven by eruptive phenomena like Coronal Mass Ejections (CMEs) and solar wind Stream Interaction Regions (SIRs)~\citep{Burlaga1971, Richardson2018,Webb2012}. IP shocks are much less investigated than Earth's bow shock due several to observational challenges, due to their higher speed with respect to the Earth's bow shock posing a stronger constraint on needed time resolution to resolve the shock transition, and due to the lower number of multi-spacecraft observations~\cite{Cohen2019}. Therefore, IP shocks allow us to access a poorly explored regime of shock dynamics, including shock evolution from their origin at the Sun and into the interplanetary medium~\citep[][]{Richardson2011}. IP shocks are typically weaker and with larger radii of curvature compared to the Earth's bow shock~\citep[e.g.,][]{Reames1999}, and their waves and particle foreshocks are much less well-characterised than their terrestrial counterparts, with several studies highlighting fundamental differences between them. For example, using the Solar Terrestrial Relations Observatory (STEREO) mission~\citep{Kaiser2008}, \citet{Kajdic2012} has shown that upstream waves are somewhat irregular at IP shocks, and are sometimes observed without corresponding shock-reflected particle populations as would be expected. \citet{BlancoCano2016} surveyed IP shocks observed by STEREO from 2007 to 2010 and showed that significant suprathermal particle populations are more likely to be found at CME-driven shocks at 1 AU with respect to SIR-driven ones due to different evolutionary features of such structures. Transient structures, routinely observed at Earth's bow shock and known to play a fundamental role in energy conversion and particle acceleration~\citep{Plaschke2018}, are very rarely observed at IP shocks, with little evidence of upstream shocklets~\citep{Lucek1997,Wilson2009, Trotta2023a} and downstream jets~\citep{Hietala2024}.

\begin{figure}[h!]
    \centering
    \includegraphics[width=.48\textwidth]{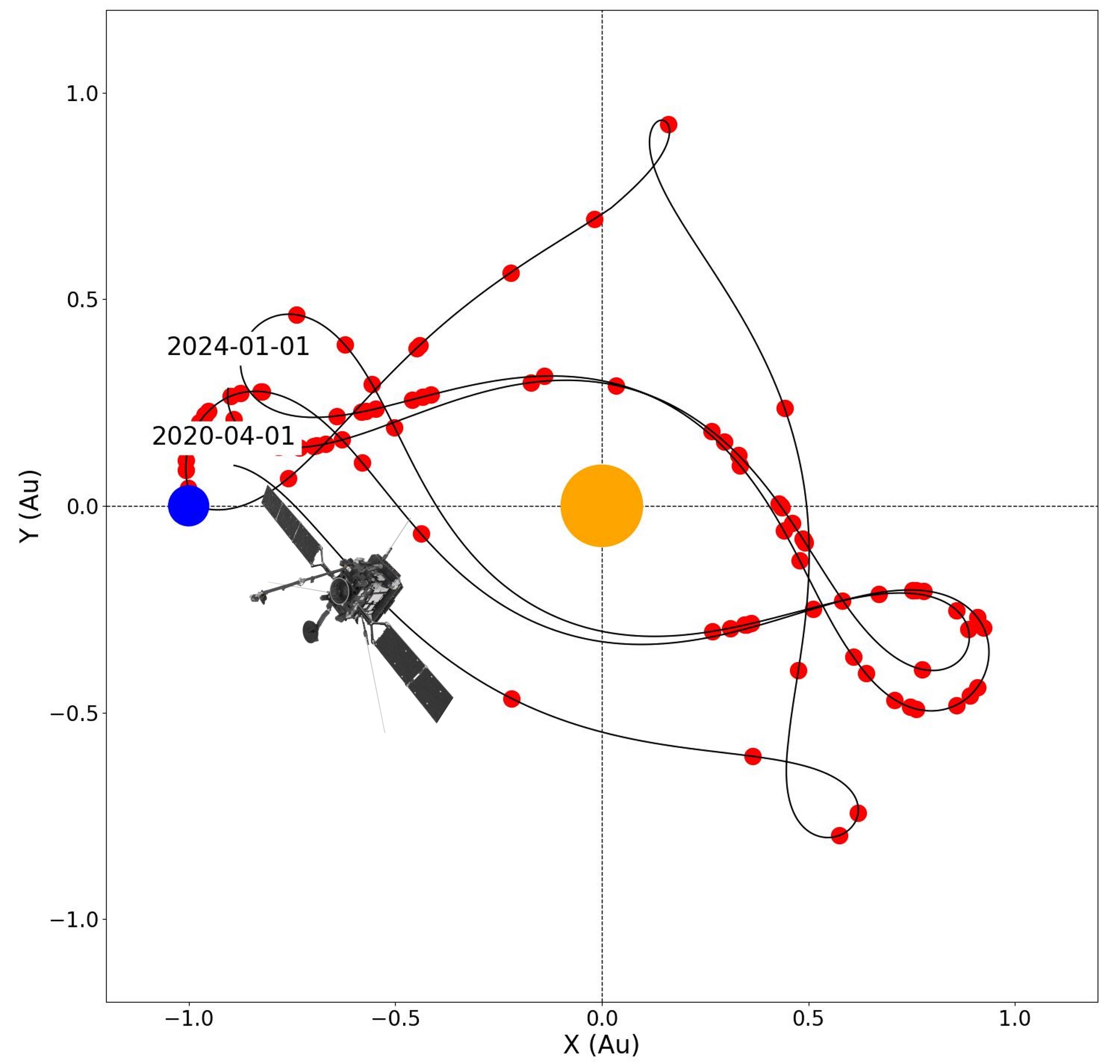}
    \caption{Solar Orbiter trajectory (black line) throughout our statistical campaign in a fixed Earth--Sun frame. The red dots represent IP shock crossings, and the blue and yellow dot represent the Earth and the Sun, respectively. (Solar Orbiter model: esa.com).}
    \label{fig:solo_traj}
\end{figure}
Additionally, IP shocks provide insights about how the shock system evolves in time and through its spatial propagation, an aspect that cannot be investigated for Earth's bow shock. From this point of view, multi-spacecraft observations leveraging different heliospheric vantage points are crucial to reconstruct fundamental properties of the shock's system~\citep{Lugaz2024}. Crucially, novel missions like NASA's Parker Solar Probe~\citep[PSP;][]{Fox2016} and ESA's Solar Orbiter~\citep{Muller2020} are probing the inner heliosphere with state-of-the-art instrumentation, thereby yielding previously unavailable datasets and therefore provide an unprecedented opportunity for discovery in IP shocks. The importance of such inner heliospheric observers has been highlighted by several recent works exploiting useful line-ups among them and the existing near-Earth spacecraft fleet~\citep[e.g.,][]{Trotta2024,Trotta2024b, Davies2024}. Furthermore, the exploitation of this tantalizing observational window is particularly timely given the  peak of activity of solar cycle 25, modulating IP shocks occurrence~\citep{Oh2007} and motivating shock surveying efforts with both long- and short-term impacts~\citep[see, for example,][]{Oliveira2023}.

\begin{table*}[h!]
 \begin{center}
 \begin{tabular}{|l|c|c|}
 \hline
 \textbf{Parameter} & \textbf{Content} & \textbf{Units} \\
 \hline
 \hline
 Shock ID & Shock unique identifier & - \\
 Shock Date & Date of shock observation & UT \\
 Shock Time & Time of the shock crossing & UT \\
 Heliocentric distance & Distance from the Sun & AU  \\
 Solar-MACH config & Link to orbit configuration plot & -  \\
 Associated SEP event & ID of associated SEP event & - \\
  Mean upstream $\delta B/B_0$  & $\delta B/B_0$ (lag: 1 min)& -  \\
 Upstream density & Mean upstream ion density & $\mathrm{cm^{-3}}$ \\
 Upstream $\beta$ & Mean upstream plasma beta & -  \\
 Upstream $\mathbf{B}$ & Mean upstream $\mathbf{B}$ vector & $\mathrm{nT}$ \\
 $\rm\hat{n}_{\mathrm{Shock}}$ & Mean shock normal vector & - \\
 $\rm\theta_{Bn}$ & Mean shock normal angle & $^\circ$   \\
 $\rm V_{\mathrm{Shock}}$ & Mean shock speed & $\mathrm{km/s}$  \\
 $\mathrm{M_A}$ & Alfv\'enic Mach number & -   \\
 $\mathrm{M_{fms}}$ & Fast Magnetosonic Mach number & -   \\
 $\mathrm{r_B}$ & Mean magnetic compression ratio & -   \\
 $\mathrm{r_{Gas}}$ & Mean gas compression ratio  & -   \\
 Structures - two hours & Structuring across shock & -  \\
 Structures - 8 minutes & Structuring across shock & -  \\
 Notes & E.g., data availability & -   \\
 Identified with & E.g., TRUFLS, Visual inspection & -   \\
 Possibility of multi-SC observation & Other SC within 0.2 AU  & -  \\
 Wave foreshock & Present/Absent & -\\
 Foreshock extent & Duration of wave foreshock & minutes  \\
 Frequency range  & Frequencies of enhanced wave activity & $\mathrm{Hz}$ \\
 Low energy particle reflection & Present/Absent  & -  \\
 Proton response at selected energies & No response/Spike/Plateau/Irregular& - \\
 Proton peak delay $\mathrm{t_{peak}} -\mathrm{t_{shock}} $ & For selected energies & minutes \\
 Electron response at selected energies & - & - \\
 Notes on particle response &- & - \\
\hline
 \end{tabular}
 \end{center}
   \caption{Summary of the shock properties and parameters provided in the Solar Orbiter shock 
   list. An interactive and downloadable list can be found on the SERPENTINE data center (see 
   Appendix~\ref{appendix:serpentine_list}).\label{tab:table_list}}
 \end{table*}

In this work, we present an extensive survey of IP shocks observed at different heliocentric distances using Solar Orbiter. The shock survey, has been carried out within the framework of the European Project Solar energetic particle analysis platform for the inner heliosphere (SERPENTINE\footnote{\url{https://serpentine-h2020.eu/}}), and is publicly available through the project data centre\footnote{\url{https://data.serpentine-h2020.eu/catalogs/shock-sc25/}}. The version discussed here is citable through Zenodo~\citep{Trotta2024CatalogueZenodo}. In Section~\ref{sec:data} we present the datasets used. Section~\ref{sec:shock_list} describes the methods of shock identification and characterisation used throughout the survey, discussing the key parameters computed for each event and summarizing the information provided for each observed IP shock. In Section~\ref{sec:results} we describe the first results of this statistical effort, showing the general trends of shock parameters with heliocentric distance and in relation to previous observations at 1~au (Section~\ref{subsec:res_oview}). We also discuss the general properties of wave foreshocks and the energetic particle response to the passage of IP shocks as observed by the novel Solar Orbiter payload (Sections~\ref{subsec:waves},~\ref{subsec:ptcls}, respectively). In Section~\ref{sec:conclusions} we present the conclusions. Finally, in Appendices~\ref{appendix:TRUFLS} and ~\ref{appendix:serpentine_list} we give details about shock search software developed and used here and about how to use this catalogue and its future implementation through the SERPENTINE data centre.

\section{Data} \label{sec:data}
We exploit the full in-situ suite on board the Solar Orbiter mission~\citep{Muller2020}. To measure the magnetic field, we use the flux-gate magnetometer~\citep[MAG;][]{Horbury2020}, yielding magnetic field measurements at up to 64 vectors s$^{-1}$. Low--energy ion energy flux has been measured with the Proton Alpha Sensor (PAS) of the Solar Wind Analyser suite~\citep[SWA;][]{Owen2020}, yielding measurements at 4s resolution in the 200 eV -- 20 keV range. For ion bulk flow speed, density and temperature we used the PAS ground moments at 4s resolution. In some cases, density estimates from the Radio and Plasma Waves instrument~\citep[RPW;][]{Maksimovic2020} are also used. Several sensors of the Energetic Particle Detector suite~\citep[EPD;][]{RodriguezPacheco2020} have been used: the SupraThermal Electons and Protons (STEP) sensor, Electron Proton Telescope (EPT) and High Energy Telescope (HET), continuously measuring energetic electrons and protons from the suprathermal regime (2 keV for electrons) to low--energy galactic cosmic ray energies of ~ 100 MeV/n at high time resolutions, up to 1s. EPT and HET comprise four telescopes with different look directions, Sunward, Anti-Sunward, North and South, respectively. The STEP detector, instead, comprises 15 pixels and its look direction overlaps with the EPT-HET Sun viewing direction.

\section{The Solar Orbiter shock list}\label{sec:shock_list}

In this Section, we describe the process of shock identification and characterisation performed to compile the Solar Orbiter shock list. Our search starts shortly after mission launch in spring 2020 and is now up to date until 31 December 2023 and containing 100 events at heliocentric distances between 0.3 and 1~au.  Figure~\ref{fig:solo_traj} shows an overview of the Solar Orbiter trajectory from April 2020 to January 2024, in a frame of reference where the Earth and the Sun are fixed (blue and yellow circles, respectively). The red dots represent the events identified, elucidating how small heliocentric distances are covered, an important complement to knowledge built on events at 1~au.

\subsection{Shock Identification} \label{subsec:identification}

Shocks can be viewed as discontinuities and, therefore, sudden changes in the plasma conditions. In this picture, identifying shocks and characterising their statistical properties in the large datasets provided by modern spacecraft missions becomes a non-trivial task, because the sudden changes in the plasma properties mentioned above occur on extremely short ($\sim$ seconds) timescales. 

Motivated by this fascinating ``big data problem'' of finding small-scale ($\sim$ seconds) structures in long time series ($\sim$ years), and in the spirit of the SERPENTINE project representing a platform for the community to study energetic particles and their origins, we decided to develop software to identify such shock transitions in spacecraft data. This software package, Tracking and Recognition of Universally Formed Large-scale Shocks (TRUFLS), is publicly available at \url{https://github.com/trottadom/PyTRUFLS} and described in detail in Appendix~\ref{appendix:TRUFLS}. The majority of Solar Orbiter shocks in this work were identified with TRUFLS, which is designed to work as well with other missions yielding in-situ magnetic field and plasma data. Once candidates are identified and confirmed visually, a full characterisation of the event is performed, as discussed in Section~\ref{subsec:estimation_techniques}. Waves and energetic particles are also thoroughly characterised, as reported in Sections~\ref{subsec:waves} and~\ref{subsec:ptcls}, respectively.

Other means of identification are used to ensure the list is complete and up to date. To this end, magnetic field-only candidates, particularly in the early stages after the launch of Solar Orbiter, when SWA suffered frequent interruptions in its operations, we often spotted by visual inspection and flagged in the list as ``magnetic field only''. Furthermore, we regularly cross-checked our shock list with the ICME catalogue ``ICMECAT'' built within the HELIO4CAST project~\citep{Mostl2017}, accessible at \url{https://helioforecast.space/}. 


\subsection{Shock parameter estimation techniques}
\label{subsec:estimation_techniques}

The structure and behaviour of collisionless shocks is regulated by several parameters, one of the most important being the angle between the normal of the shock surface and the upstream magnetic field \tbn~\citep[e.g.,][]{Burgess2015}. For \tbn{} values close to 90$^\circ$, i.e, when the upstream magnetic field is almost tangential to the shock surface, the shock is quasi-perpendicular. On the other hand, for \tbn{} values close to 0$^\circ$ (corresponding to an upstream magnetic field almost normal to the shock surface), the shock is quasi-parallel. Particle reflection and propagation far upstream is favoured at quasi-parallel shocks~\citep[][]{Kennel1985}, introducing the possibility for reflected particles to interact with the upstream plasma over long distances, creating unstable distributions and a collection of disturbances in the plasma properties, giving rise to the so-called particle and wave foreshocks~\citep[e.g.,][]{Eastwood2005}.

Other important parameters that dictate the behaviour of collisionless shocks are the (Alfv\'enic and fast magnetosonic) Mach number, i.e., the ratio between the shock speed in the upstream flow frame ($\rm v_{sh}$) and the upstream Alfv\'en ($\rm v_{A}$) and fast magnetosonic ($\rm v_{fms}$) speeds, respectively ($\mathrm{M_A \equiv v_{sh}/v_A}$ and $\mathrm{ M_{fms} \equiv v_{sh}/v_{fms}}$). The upstream plasma beta, i.e., the ratio between the plasma and magnetic field pressure, often expressed as a ratio of squared thermal and Alfv\'en speeds $\rm\beta \equiv v_{th}^2/ v_{A}^2$, and the gas and magnetic compression ratios ($\mathrm{r_{Gas} \equiv n_d/n_u}$ and $\mathrm{r_B \equiv B_d/B_u}$, respectively, where $d$ and u subscripts indicate the downstream and upstream states) are also relevant for shock dynamics.

Shock parameter estimation using single spacecraft crossings is often a challenging task due to the intrinsic three-dimensional nature of the system and the fluctuations typically involved in the transitions~\cite [e.g.][]{Koval2008}. Throughout this work, we used different methods to determine shock parameters. The shock normal (and therefore the \tbn) estimation is done using the Mixed Mode 3 method (MX3), consistent with previous catalogs~\citep{Kilpua2015_shocks}. When plasma data is not available, the magnetic coplanarity (MC) method is used to determine the shock normal vector. The shock speed along its normal is computed using the mass flux conservation, and it is given in the spacecraft frame of reference. A comprehensive summary of the above techniques can be found in \citet{Paschmann2000}. From these estimations, further shock parameters are determined (e.g., Mach numbers, compression ratios), as summarized in Table~\ref{tab:table_list}.

Crucially, single spacecraft shock parameter estimations involve an operation of averaging upstream/downstream of the shock crossing, making the results particularly sensitive to the choice of averaging windows. Furthermore, the shock normal determination is particularly sensitive to the choice of averaging, introducing an uncertainty that propagates to shock speed and Mach number estimations carried out in the shock normal frame~\citep{Paschmann2000}. Building on the idea first proposed in~\citet{Balogh1995}, we developed a technique involving a systematic variation of upstream/downstream averaging windows and yielding distributions of shock parameters which can be used to determine uncertainties~\citep[see][for details]{Trotta2022b}. Shock parameters in this catalogue have been computed with such a technique, implemented within the SERPENTINE project in the publicly available SerPyShock code\footnote{\url{https://github.com/trottadom/SerPyShock}}. The length of the averaging windows was systematically varied from 30 seconds to 8 minutes both upstream and downstream, compatible with fixed windows previously used in other catalogues~\citep{Kilpua2015_shocks}. 

\begin{figure*}[]
\centering
    \includegraphics[width=.99\textwidth]{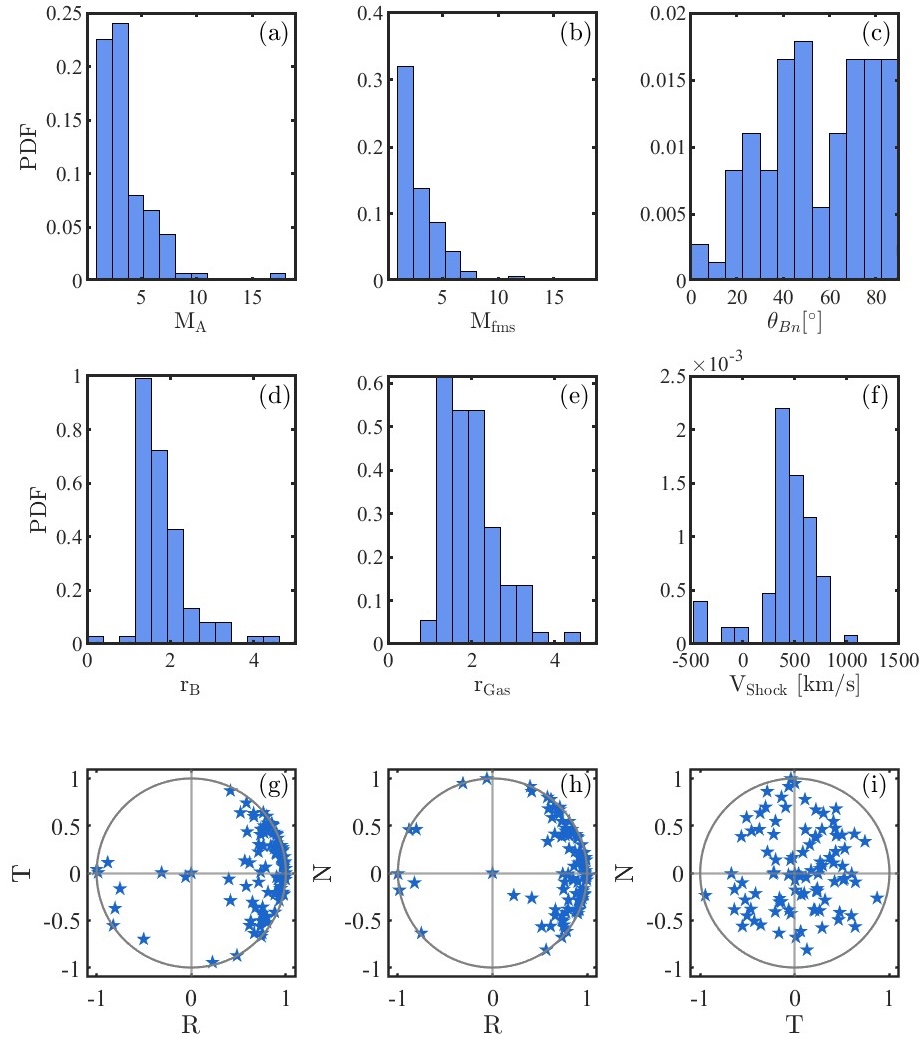}
    \caption{Overview of shock parameters for all the shocks identified. The panels show PDFs of Alfv\'enic and fast magnetosonic Mach numbers 
    $\mathrm{M_A}$, $\mathrm{M_{fms}}$, shock normal angle \tbn, magnetic and gas 
    compression ratios $\mathrm{r_B}$, $\mathrm{r_{Gas}}$, and shock speed $\mathrm{V_{Shock}}$ (a-g, 
    respectively). Panels g-i show scatters of shock normal vectors in the RTN frame.  \label{fig:histograms}}
\end{figure*}

The Solar Orbiter shock list is publicly available on the SERPENTINE data centre, where interactive access and a series of quicklook plots are also included. These are detailed in Appendix~\ref{appendix:serpentine_list}. In Table~\ref{tab:table_list}, we summarised the quantities reported in the list and used in the present work. It may be noted that the structure of the list can be divided into three parts: identification and context, shock parameters, and wave and energetic particle response (top to bottom in Table~\ref{tab:table_list}).

The link to the Solar-MACH configuration plot (entry 5 in Table~\ref{tab:table_list}, available in the online version of the catalogue) uses the Solar MAgnetic Connection HAUS tool~\citep{Gieseler2023}, yielding the orbital configuration for the event. Additionally, we provide information regarding association with SEP events referring to the  Solar Cycle 25 multi-spacecraft SEP event catalogue of the SERPENTINE project~\citep{Dresing2024}, also available through the SERPENTINE data centre\footnote{\url{https://data.serpentine-h2020.eu/catalogs/sep-sc25/}}. The mean upstream $\delta B/B_0$, density, plasma $\beta$ and magnetic fields are evaluated in the 8 minutes before the shock. This choice is fixed for all events, to ensure reproducibility and the possibility to seamlessly integrate new events in the future. A case-by-case procedure, based on the visual inspection of each event is also reported in Appendix~\ref{appendix:dimmock_list}.

\section{Results}\label{sec:results}

In this Section, we show the outcomes of the identification and characterisation described in Section~\ref{sec:shock_list} for the first 100 Solar Orbiter shocks detected from 2020 to 2023 at heliospheric distances spanning 0.3 -- 1~au.

\begin{figure*}[]
\centering
    \includegraphics[width=.99\textwidth]{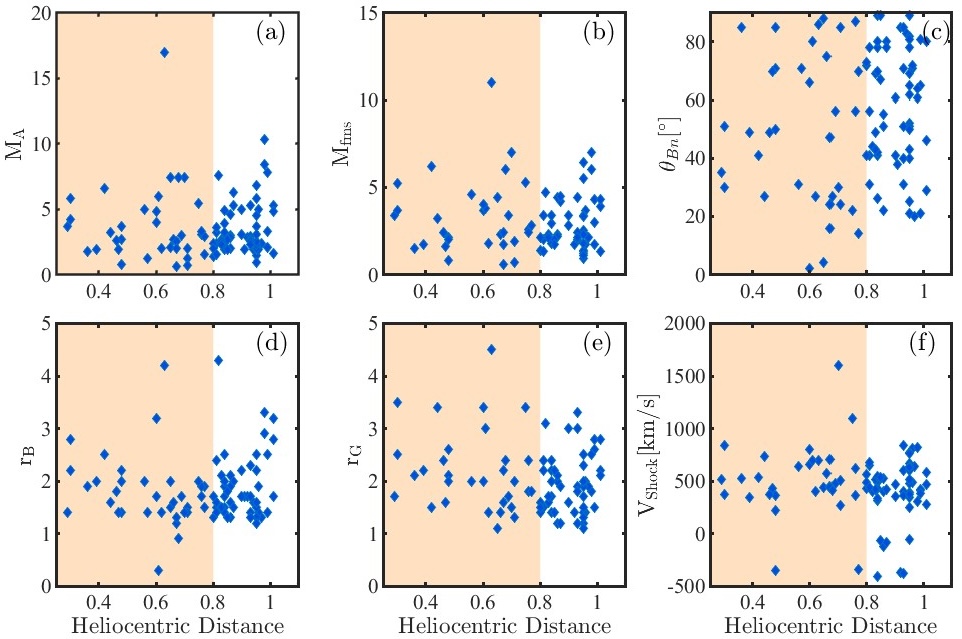}
    \caption{Shock parameters as a function of heliocentric distances in au. The parameters shown in panels a-f follow from 
    the same panels in Figure~\ref{fig:histograms}. The orange-shaded panels highlight poorly investigated 
    heliocentric distances.   \label{fig:rad_trends}}
\end{figure*}

\begin{figure*}[]
\centering
    \includegraphics[width=.99\textwidth]{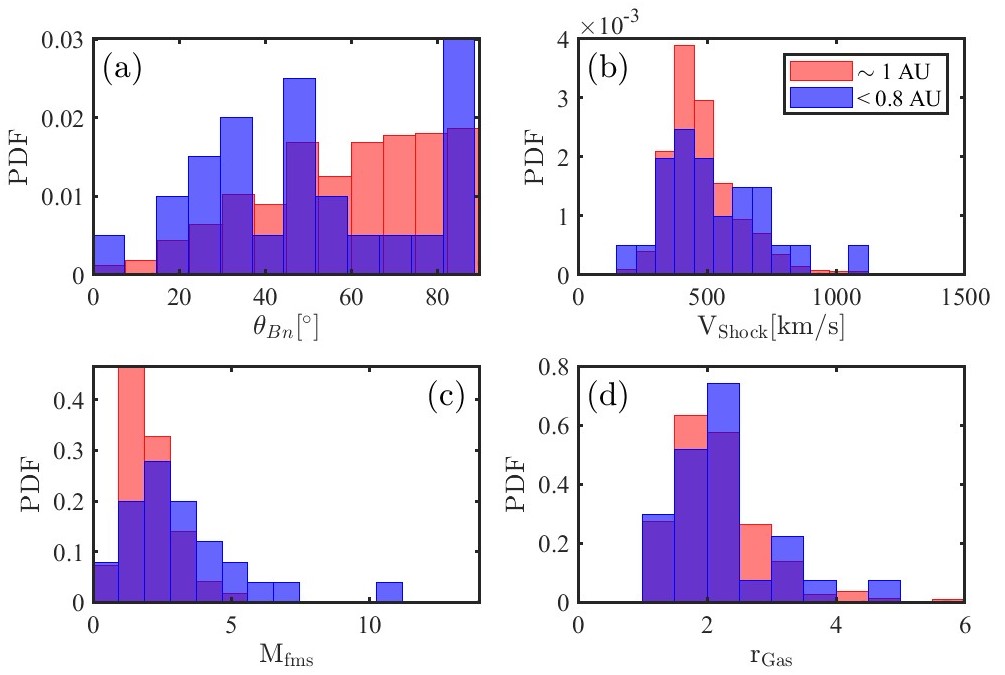}
    \caption{Comparison of shock parameters as observed by Solar Orbiter for 30 shocks at heliocentric distances smaller than 0.8 AU (blue histograms) and 586 shocks observed from 1995 to 2009 at 1 AU by~\citet{Kilpua2015_shocks}. \label{fig:hist_compare_1au}}
\end{figure*}

\subsection{Overview}\label{subsec:res_oview}

Figure~\ref{fig:histograms} shows an overview of shock parameters for all the 100 events in the list. Typical features of IP shocks are recovered here, as described below.  Alfv\'enic and fast magnetosonic Mach numbers are mostly moderate ($\leq 3$), confirming that IP shocks tend to be weaker than the Earth's bow shock~\citep[e.g.,][]{Lalti2022}. Both gas and magnetic field compression ratios are systematically lower than 4~\citep[e.g.][]{SMith1985, Lario2005, Kilpua2015_shocks, PerezAlanis2023}. The distribution of shock normal angles \tbn{} (Figure~\ref{fig:histograms}c) shows that the quasi-perpendicular geometry (\tbn $\geq$ 45$^\circ$) is favoured for IP shocks. This effect is probably due to two factors, namely the global configuration of the magnetic field the shocks propagate through~\citep{Chao1985,Reames1999,Janvier2014}, and a selection bias because quasi-parallel shocks are more challenging to identify~\citep{Kruparova2013,Oliveira2023}. In Figure~\ref{fig:histograms}f, we show the distribution of shock speeds. In this plot, shocks with negative speed are fast reverse shocks, which rarely occur in the inner heliosphere with respect to 1~au and beyond~\citep{Schwenn1996, Jian2006}. However, given the high level of solar activity, some  reverse shocks may be observed, for example due to CME--CME interactions, as recently shown in~\citet{Trotta2024b}. Finally, in Figure~\ref{fig:histograms}g-i we show the orientations of the local shock normals, in the RTN frame of reference, for all the events. It may be noted that the local shock normal vectors exhibit a strong variability, indicating that  strong departures from the radial directions are routinely observed locally, thus highlighting the variability of the shock system.

In Figure~\ref{fig:rad_trends}, we further exploit the Solar Orbiter capabilities in probing the inner heliosphere, showing the shock parameters for all the events as a function of heliocentric distance.  Heliocentric distances where Solar Orbiter spent  a limited time, and therefore identified only a few shocks, are highlighted by the orange shaded panels. As shown in Figure~\ref{fig:rad_trends}, no strong trends can be identified for shock parameters at different heliocentric distances, except for \tbn{} going towards more perpendicular geometries with increasing heliocentric distance, as described above. There is, however, a small indication that the three shocks identified near perihelion (0.3 AU) may be relatively strong, an effect that could be related to faster CMEs driving such shocks in the inner heliosphere~\citep[e.g.,][and references therein]{Balmaceda2020}. These results indicate that the local shock behaviour is poorly influenced by the shock age and/or different global heliospheric configurations, with important implications for energetic particle production in the heliosphere.

To take the characterization of inner heliospheric shocks further, we filtered the sample based on heliocentric distance and isolated 30 fast-forward shocks below 0.8 AU and compared them with 586 fast--forward shocks observed with STEREO-A and \textit{Wind} between 1995 and 2009, a very large sample reported in~\citet{Kilpua2015_shocks}. Results for selected shock parameters are shown in Figure~\ref{fig:hist_compare_1au}. It can be seen that there are no significant deviations from the overall behaviour at 1 AU, apart from a slight trend in \tbn (Figure~\ref{fig:hist_compare_1au}a), in which more parallel configurations are favoured at low heliocentric distances. Further, there is indication for slightly higher shock speeds and fast magnetosonic Mach numbers in the inner heliosphere, possibly due to faster shock drivers at short heliocentric distances. Such indication is compatible with what previously found on IP shock with the Helios mission~\citep{Volkmer1985,Lai2012,Hajra2023}. A more detailed investigation of radial trends with modern observatories will be the object of further work, including input from other important statistical studies using other inner heliospheric missions, such as Parker Solar Probe~\citep{Fox2016} and BepiColombo~\citep{Benkhoff2021}, allowing in particular larger statistics of shocks closest to the Sun (< 0.5 au).

\subsection{Wave environment}\label{subsec:waves}
In this section, we briefly report on the wave environment observed at Solar Orbiter shocks. A detailed characterisation of magnetic field fluctuations in correspondence to IP shock crossings has been performed for all the events in the Solar Orbiter shock list. The first quantity computed is the value of $\delta B/B_0$ close to the shock, computed as $\delta B/B_0 \equiv |\mathbf{B}(t+\tau) - \mathbf{B}(t)|/|\mathbf{B}(t)|$, where the lag $\tau$ has been chosen of 1 minute (relevant to resonant scattering of protons with energies of around 100 keV) and then averaged for 8 minutes upstream. Furthermore, for each event we study wavelet spectrograms of magnetic field intensity $B$ and trace spectrograms of the magnetic field components. From the latter, the duration of the foreshock is estimated by visual inspection of intensity enhancement in correspondence with the crossing, specifying the range of wavelengths where such enhanced fluctuations are observed (see Table~\ref{tab:table_list}). Finally, to gain insights about the polarisation of the observed waves, the reduced magnetic helicity, normalised by the power in magnetic field fluctuations $\rm\sigma_{m}(k) \equiv k H_m^{(r)}(k)/E_B(k)$ (where $\rm k$ is the wavenumber,  $\rm H_m^{(r)} (k)$ is the reduced magnetic helicity \citep{Matthaeus1982} and $\rm E_B(k)$ is the magnetic power spectral density) is also studied~\citep{Bowen2020,Woodham2021}. 
Wave foreshocks result from the interaction of shock--reflected particles and the upstream plasma, where unstable distributions give rise to magnetic fluctuations extending far into the shock upstream, crucial for particle acceleration~\citep[e.g,][]{Burgess2015}. Another type of upstream waves, known as precursors, is often observed but generated by different shock physics. These waves are commonly observed between around 0.5Hz - 10 Hz, primarily classified as whistler waves, a branch of the fast magnetosonic wave. Whistler waves are also reported in the hundreds of Hz range but are not considered in these statistics. However, evidence of them has been observed in a few Solar Orbiter IP shocks (not shown).

Whistler waves can also be affected by Doppler-shift \citep{dimmock2013}, resulting in some variability in their frequency in the spacecraft frame. Whistler waves are a vital mechanism that aids the balancing of the nonlinear steepening of the ramp, predominantly observed around quasi-perpendicular shocks, and may be crucial for the first steps of electron acceleration~\citep{Riquelme2011}. There are still many open questions about their generation and  role in energy conversion at the shock front. Since we do not identify the wave mode here, we will refer to these waves as low-frequency precursors. Considerable publications have examined these precursors using various techniques and instruments, confirming they are both a critical component of the shock magnetic structure and intrinsically connected to particle dynamics across the shock. As expected, most analyses have benefited from decades of multi-spacecraft observations collected at the terrestrial bow shock \citep{Fairfield1974,dimmock2013,lalti2021}, leading to many noteworthy results that have significantly advanced our understanding of the role such waves play in collisionless shock dynamics. There have also been investigations of low-frequency precursors at different planets such as Venus \citep{dimmock2022}, Mars \citep{brain2002}, Saturn \citep{sulaiman2016}, Mercury \citep{fairfield1976}. Moreover, they are also repeatedly observed in the solar wind close to interplanetary shocks \citep{Wilson2009}, which is the focus of this paper. Our goal is to determine how many shocks in the Solar Orbiter database contain low-frequency precursors. To decide if low-frequency precursors are present, we inspect the wavelet power and ellipticity to check for enhancements and circular polarisation in the relevant frequency range. The shocks for which no MAG burst or SWA-PAS data was available, were excluded, resulting in 72 shocks to calculate this statistic. Based on the criteria above, visual assessment determines if a low-frequency precursor train is present. 

Finally, we investigate, by visual inspection, the presence of pre-existing structures such as flux ropes and discontinuities. Such structures, often neglected in theoretical modelling of particle acceleration, are emerging as a fundamental ingredient for particle acceleration~\citep{Guo2021}. Indeed, the interaction of such pre-existing structures with shocks may lead to enhanced  energetic particle production, ranging from enhanced scattering upstream to downstream trapping~\citep[][]{Giacalone2021,Trotta2020b,Kilpua2023}. In the catalog, the presence of such structures is reported for intervals in the 8 minutes and 2 hours around the shock (see Table~\ref{tab:table_list}).

\begin{figure*}[]
\centering
    \includegraphics[width=.99\textwidth]{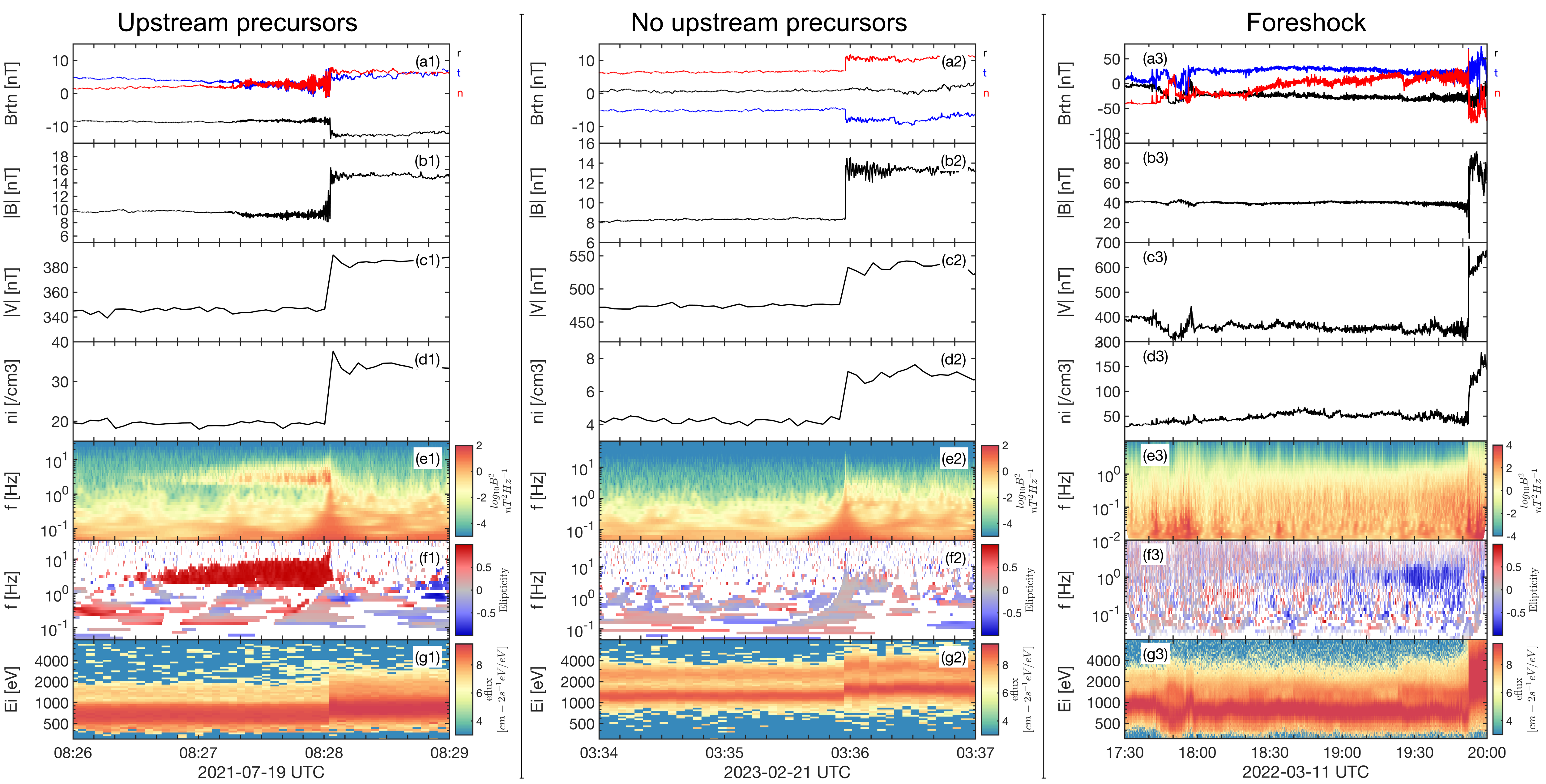}
    \caption{Examples of three interplanetary shocks observed by Solar Orbiter exhibiting different wave environments. Panels (a-g) show $\mathbf{B}$, $|\mathbf{B}|$, $\mathbf{V}$, $n_i$, wavelet spectra, magnetic field ellipticity, and omnidirectional energy flux. The first case shows a clear upstream precursor (a1-g1), the middle one (a2-g2) shows no wave activity and finally the case on the right has an extended wave foreshock (a3-g3).   
  \label{fig:precursors}}
\end{figure*}

Wave foreshocks have been identified in the frequency range between 0.01 and 1 Hz for 46\% of the shocks. The duration of foreshocks has been found to vary from a few minutes to about one hour, depending on the pre-existing structuring of the medium  the shock propagates through. Indeed, long wave foreshocks have been found in cases where the shock propagates in plasma exhibit low levels of pre-existing magnetic fluctuations. One example is the long wave foreshock observed within CME material and documented in~\citet{Trotta2024b}. Extended wave foreshocks in such ``quiet'' environments are compatible with observations of long-lasting field-aligned beams of energetic particles, where scattering is inhibited under similar local plasma conditions~\citep{Lario2022}.
Similarly, we estimate that wave precursors are present in approximately 62\% of shocks in this database, with the intriguing property of showing wave trains with significant changes in their duration. Solar Orbiter observations confirm that IP foreshocks tend to occur for low \tbn{} and high Mach numbers, a topic that will be expanded in further studies.


\begin{figure*}[]
\centering
    \includegraphics[width=.99\textwidth]{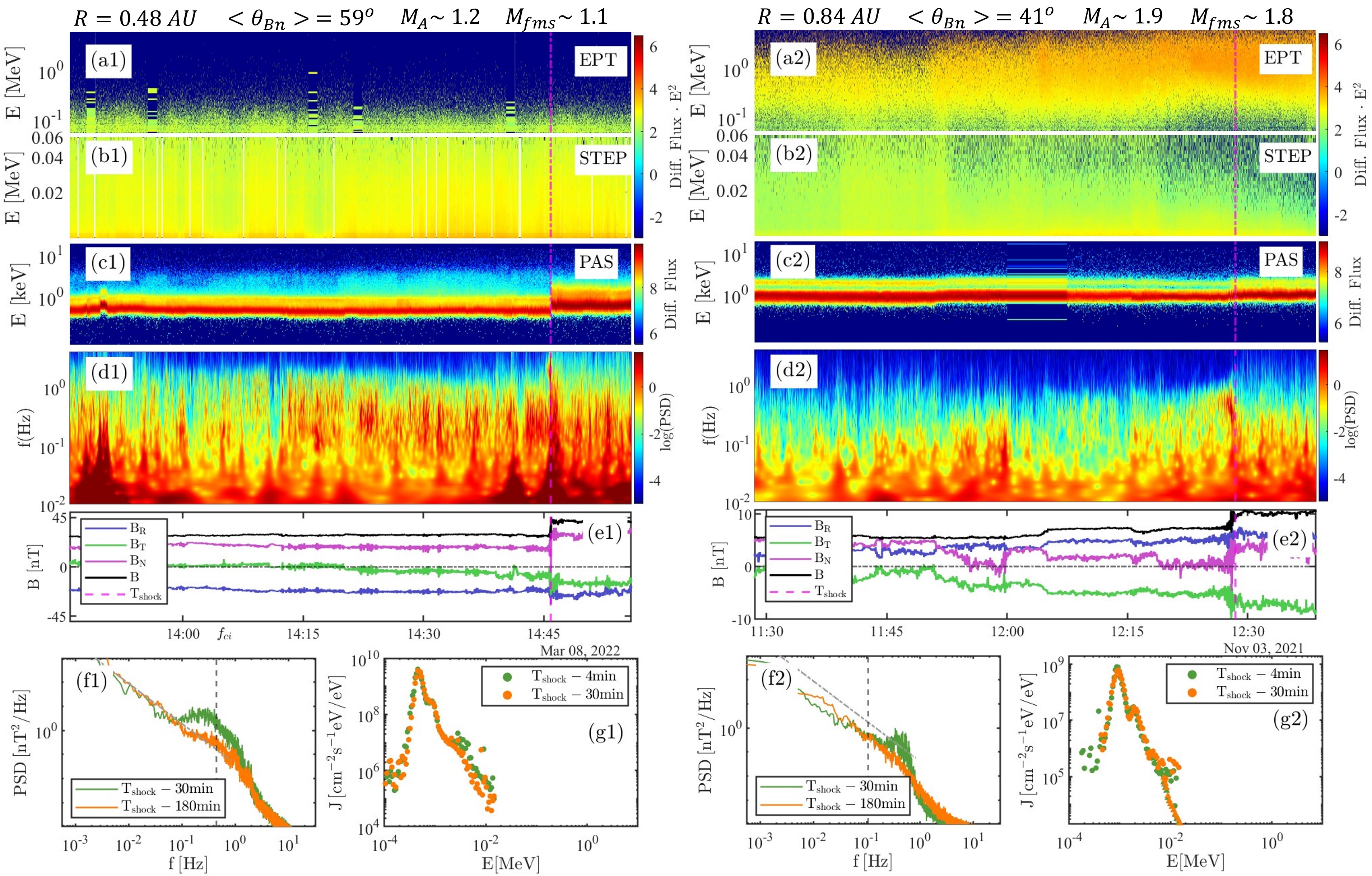}
    \caption{Wave and particles foreshocks for two different events. a--b): Energetic differential fluxes (in $\mathrm{E^2 \cdot cm^{-2} s^{-1} sr^{-1}~MeV}$) as measured by EPD's Sun-directed EPT sensor (a) and STEP sensor (b). c):  Energy flux (in $\mathrm{cm^{-2} s^{-1}~eV}$) measured by SWA-PAS. d-e): Magnetic field trace wavelet spectrogram (d) and magnetic field magnitude and components (e) measured by MAG. f) Magnetic field PSD collected far (orange) and close (green) upstream of the shock. g) PAS-EPD Ion energy spectra were collected far (orange) and close (green) upstream of the shock. Example 1 shows both a wave and particle foreshock, while Example 2 shows a significant wave response but no particle counterpart.\label{fig:waves_nopart}}
\end{figure*}

Figure~\ref{fig:precursors} shows three examples of shocks with different wave environments. From top to bottom, we show the magnetic field vector components in spacecraft-centered RTN coordinates and magnitude, ion bulk flow speed and density, the trace wavelet power spectral density PSD spectrum of the magnetic field, the reduced magnetic helicity and the one--dimensional energy flux measured by PAS in a range of time scales ranging from a few minutes to about 3 hours. The case on the left column shows a shock with visible precursors, as shown in panels (e1) and (f1), where enhanced wavelet power and the elliptical polarization are evident in the $\sim$ 2 minutes interval preceding the shock crossing. In panel (f1), a red cluster indicates the circularly right-handed polarization. Their frequency in the spacecraft frame is around 3 Hz, and they are visible for up to 90 seconds upstream. This shock has $\theta_{Bn} = 67^\circ$, $M_A = 2.8$ and was observed at 0.8 AU. In contrast, the shock in the middle column  of Figure~\ref{fig:precursors} (a2-g2) has no clear indications of upstream precursors and any other wave activity upstream. This shock was observed around 0.8 AU and had $M_A=2.4$, similar to the first shock. However, this shock was almost perpendicular with $\theta_{Bn} = 86^\circ$. Another difference is that the second shock appears to have additional fluctuations downstream of the shock front compared to the first. Finally, the right side coloumn of Figure~\ref{fig:precursors} shows an interplanetary shock with a wave foreshock lasting about 10 minutes observed at 0.44~au. This shock is quasi-parallel ($\theta_{Bn}~27^\circ$) and characterised by higher Mach numbers ($\rm M_A \sim M_{fms} \sim 3.2$). The fluctuations enhancement is well-visible in the wavelet spectrogram at frequencies between 0.01 and 1 Hz. 

\begin{figure*}[]
\centering
    \includegraphics[width=.99\textwidth]{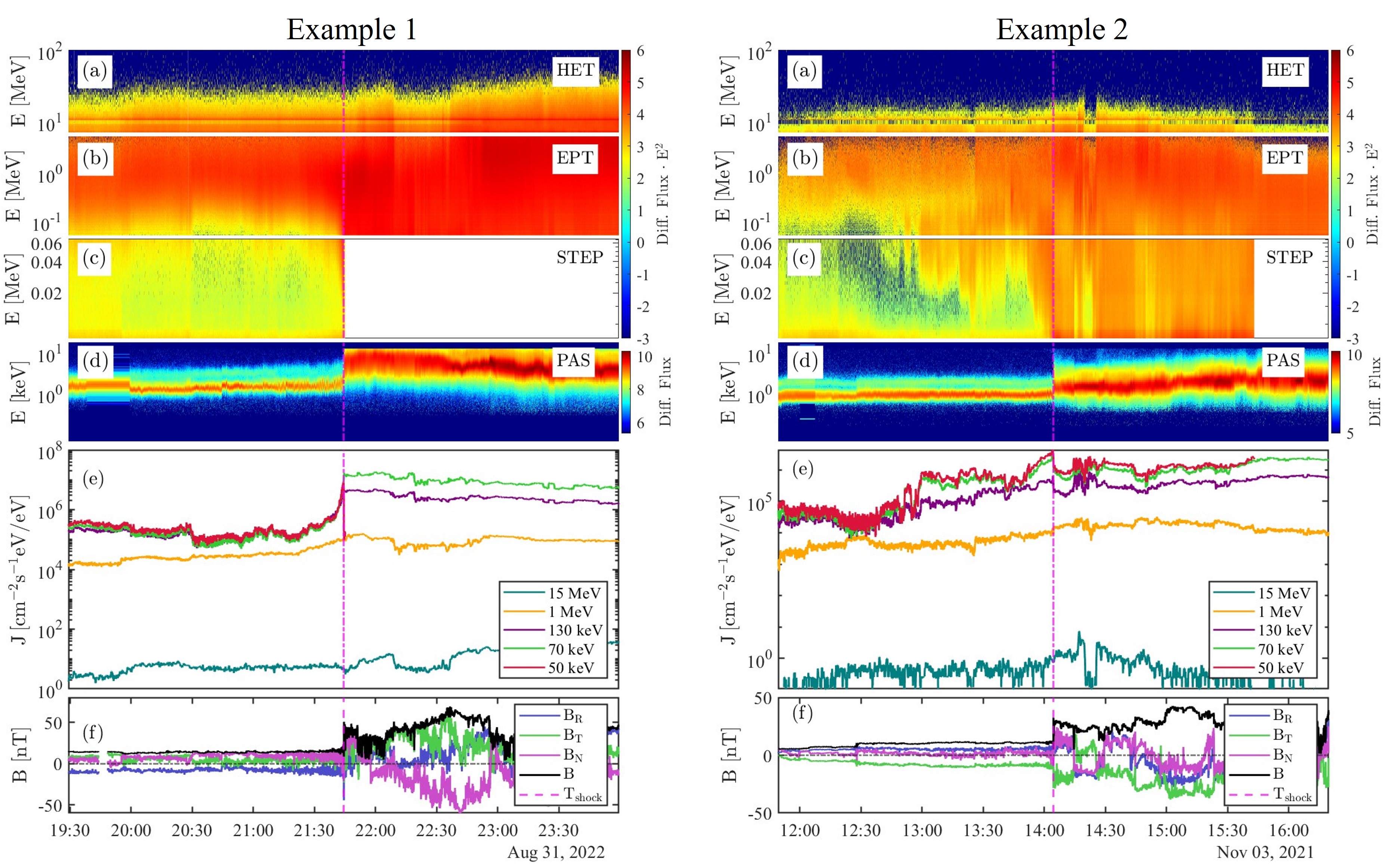}
    \caption{Two examples of energetic protons Solar Orbiter observations in response to an IP shock passage. a-c) Spectrograms of energetic ions differential fluxes (in $\mathrm{E^2 \cdot cm^{-2} s^{-1} sr^{-1}~MeV}$) as measured by EPD's Sun-directed HET (a) and EPT (b) sensors, and by the STEP sensor (c). d):  Energy flux (in $\mathrm{cm^{-2} s^{-1}~eV}$) measured by SWA-PAS. e) Ion differential fluxes for selected energy channels. f): Magnetic field magnitude and components. \label{fig:examples_particles}}
\end{figure*}

Figure~\ref{fig:precursors} shows how the variety in this database will help advance our understanding of low-frequency precursors' appearance and their role in interplanetary shocks. It will also allow IP shocks to be easily compared to databases collected in various other plasma environments \citep[e.g.,][]{Lalti2022, PerezAlanis2023}.

One example of interesting comparison across different environments is the association of wave foreshocks with shock--reflected particles that can lead to unstable upstream distributions, as routinely observed at Earth's bow shock~\citep{Kucharek2004, Archer2005}. Such association is often less clear for IP shocks. In Figure~\ref{fig:waves_nopart}, we show two examples of wave and particle response to the passage of IP shocks. In the Figure, from top to bottom, we show: the EPT-Sun, STEP and PAS energy flux spectrograms (a-c), the trace magnetic field wavelet spectrogram (d), magnetic field magnitude and its RTN components (e), the magnetic field power spectral density collected in a 30 minutes window immediately upstream of the shock and 2 hours before the shock arrival (green and orange lines, f panels), and one dimensional ions energy fluxes collected for 4 and 30 minutes before the shock (green and orange points, g panels). On the left side of the Figure, we show an oblique ($\theta_{Bn} \sim 59^\circ$) shock with low Mach number preceded by an extended ($\sim$ 30 min) wave foreshock and an enhancement of superathermal particles upstream. This event happens in an environment with a low level of magnetic fluctuations typical of of CME--material and, despite the low Mach number of the shock, provides an exceptional opportunity to study the interplay between upstream waves and shock-reflected particles (as reported in ~\citet{Trotta2024b} and Blanco Cano et al., \textit{in prep.}). In this case, despite the low Mach number of the shock, we find both an usually long ($\sim$ 30 mins) wave foreshock and a population of shock reflected particles upstream (Figure~\ref{fig:waves_nopart} f1, g1). On the right side of the Figure, we show a different case of a $\theta_{Bn} \sim 41^\circ$ shock, showing enhanced wave activity close ($\sim$ 2 min) to the shock  and up to 1 Hz (see Figure~\ref{fig:waves_nopart} d2) but no suprathermal particle counterpart (Figure~\ref{fig:waves_nopart} g2). Such suprathermal counterpart was found instead in other cases with similar shock parameters case described in \citet{Dimmock2023}. Figure~\ref{fig:waves_nopart} shows the emerging complexity of IP shocks, where many effects, ranging from time evolution to spatial irregularities and subsequent spacecraft connectivity to the shock surface~\citep[e.g.,][]{Kajdic2012} led to unexpected observations based on knowledge built on the Earth's bow shock.

\begin{figure*}[]
\centering
    \includegraphics[width=.99\textwidth]{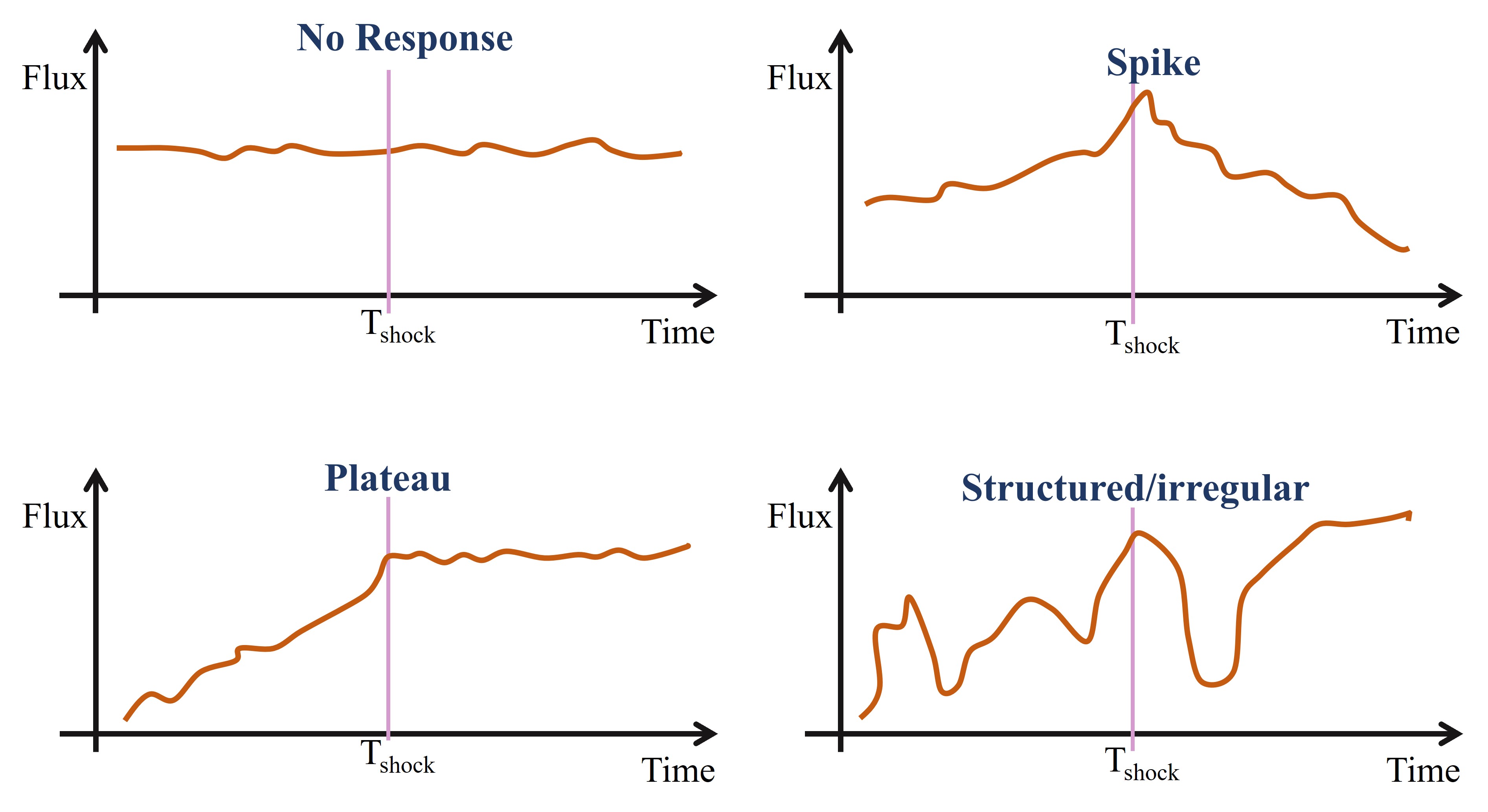}
    \caption{Sketch of the type of energetic particle flux responses to the IP shock passage.    \label{fig:responsesketch}}
\end{figure*}

\subsection{Energetic Particles}\label{subsec:ptcls}
In this Section, we report an overview of the energetic particles' behaviour observed for the IP shocks in the sample. Solar Orbiter PAS one-dimensional energy fluxes have been used to address and report the presence of reflected particles immediately upstream of the shocks (see Figure~\ref{fig:waves_nopart}), a feature that is relatively hard to resolve for IP shocks ~\citep{Dimmock2023}. At higher energies, we use the unprecedented time-energy resolution capabilities of the Solar Orbiter EPD suite to characterise the energetic particle response for each shock crossing. Most of the discussion below is centred around the response of ions to an IP shock passage. Electron response is also characterised and provided in the SERPENTINE shock list, but with less detailed information, given that IP shocks rarely accelerate electrons efficiently in-situ ~\citep{Dresing2016}. 
A very detailed and extensive statistical campaign investigating energetic particle production in relation to the ambient/shock parameters is the object of a separate investigation (Kartavykh et al., \textit{in prep.}).


Figure~\ref{fig:examples_particles} shows two examples of strong shocks associated with significant ion acceleration, where spectrograms of energetic (STEP, EPD-Sun, HET-Sun) and thermal (PAS) particle fluxes are shown (a-d) together with energetic particle flux profiles in the selected energy channels reported in the shock list (50 keV, 70 keV, 130 keV, 1 MeV, 15 MeV, panels e). Finally, the magnetic field and its components are also shown (panels f). EPD particle spectrograms (panels a-c) make the very high time-energy resolution of such datasets particularly clear.

In both cases, the shocks are propagating through an already enhanced energetic proton population, due to a previous phase of the event or to a previous SEP event. In Figure~\ref{fig:examples_particles}--Example 1, we observe energetic protons' differential fluxes up to 1 MeV rising exponentially upstream up to the shock crossing and then becoming constant downstream. Such time--energy profiles can be interpreted as a signature of Diffusive Shock Acceleration\citep[e.g.,][]{Axford1977,Blandford1978,Bell1978a} operating at the interplanetary shock. Such observation is compatible with the shock parameters measured locally, indicating a quasi-parallel ($\theta_{Bn}\sim22^\circ$), fast ($V_{sh}\sim 1100$ km/s shock with high Mach numbers ($M_A \sim M_{fms} \sim 5.4$). For 15 MeV particles, a more complex behaviour is observed, with an irregular response to the shock passage. This can be put in the context of the highly variable environment observed in the shock downstream, with magnetic structures modulating energetic particle fluxes, with a weaker effect  of the shock at affecting the behaviour of these high energies. It may be noted that the 15 MeV intensity peaks downstream of the shock, which may be an essential region for further particle acceleration, as shown as well by the recent observation of strong IP shocks~\cite [see][]{Lario2003, Kilpua2023}.

\begin{figure}
\centering
    \includegraphics[width=.49\textwidth]{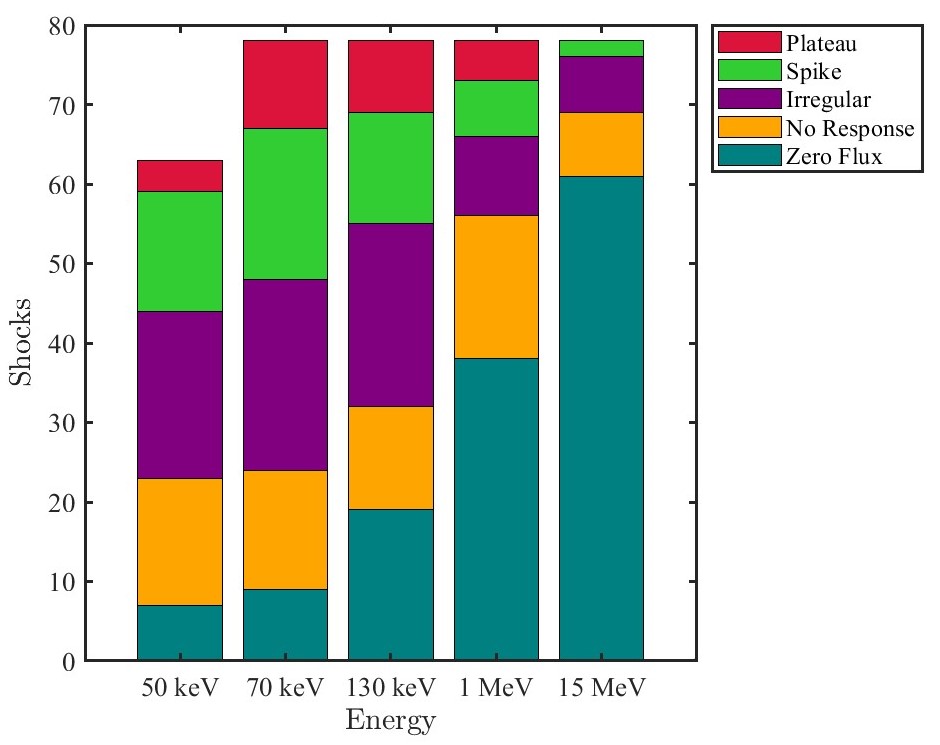}
    \caption{Bar chart describing the type of ion responses (Figure~\ref{fig:responsesketch}) at different energies for the whole IP shock sample. \label{fig:bar_response}}
\end{figure}

The theme of complexity concerning energetic particle production is even more evident in Figure~\ref{fig:examples_particles}--Example 2, where another strong shock is shown. While the standard collection of averaging windows used for the SERPENTINE shock list yields a \tbn of 51$^\circ$ and moderate Mach numbers ($M_A \sim 3 ; M_{fms} ~ 2.3$), we note that shock parameter estimation is particularly challenging in this event, due to strong disturbances in the magnetic field reported upstream/downstream of the 
shock, as reported in \citet{Trotta2023a}. Such complexity yields irregular time profiles of energetic particles, where many different mechanisms of production of suprathermal and high energy protons have been found by recent studies~\cite{Yang2023,Yang2024}. Once again, in Example 2 the particle flux at the highest energy reported (15 MeV) peaks downstream of the shock, highlighting the importance of the pre--existing features where IP shocks propagate.

As intensity--time profiles of energetic particles yield invaluable information about particle acceleration and transport at IP shocks, we characterised them for the Solar Orbiter sample. We study the energetic ions fluxes for the selected energy channels of 50 keV, 70 keV, 130 keV, 1 MeV and 15 MeV (all in the spacecraft rest frame), thus employing EPD-STEP for the 50 keV response, EPD-EPT Sun for the 70 keV, 130 keV and 1 MeV response, and EPD-HET Sun for the 15 MeV channel. The profiles are visually characterized as ``spike'', ``plateau'', ``structured/irregular'', ``no response'', summarised in the sketch reported in Figure~\ref{fig:responsesketch}. This effort continues earlier surveys of shocks at 1 AU in previous Solar Cycles ~\citep[e.g.,][]{Lario2003}, making our study particularly relevant due to the novel energy-time resolution capabilities provided by the Solar Orbiter EPD suite and the fact that a poorly investigated range of heliocentric distances can be explored.

\begin{figure*}
\centering
    \includegraphics[width=.99\textwidth]{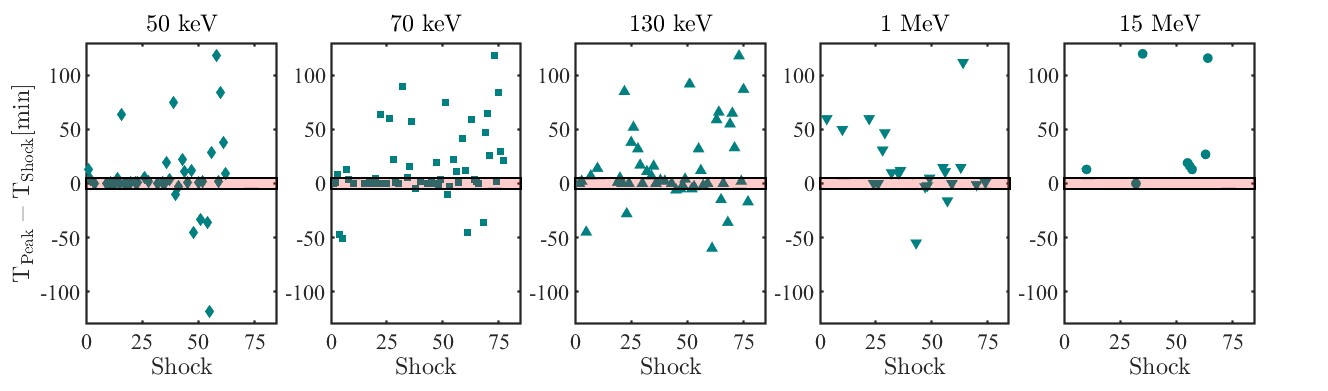}
    \caption{Time of peak vs time shock $\rm T_{Peak} - T_{Shock}$ for each event, for all the selected energy channels (left to right). The red boxes represent a 10-minute window around $\rm T_{Peak} - T_{Shock} = 0$. \label{fig:TmaxvsT}}
\end{figure*}

The outcome of our analysis is shown in Figure~\ref{fig:bar_response}, showing a bar chart characterising energetic particle response to the IP shock passage for the selected energy channels. The colors were chosen as follows: teal indicates ``Zero Flux'', meaning no energetic particles were detected in the interval surrounding the shock. Then, yellow, purple, green and red indicate the ``no response'', ``irregular'', ``spike'' and ``plateau'' responses outlined in Figure~\ref{fig:responsesketch}. The ``no response'' case is different from the ``Zero flux'' one as it has a significant flux but no changes in particles fluxes are detected in relation to the shock crossing. Note that, in Figure~\ref{fig:bar_response} the 50 keV bar is shorter since EPD-STEP data is available for fewer shocks than EPT and HET, with 64 shocks participating in the analysis for STEP and 79 for EPD and HET. In our data set, 88.5\% of the shocks do not show a significant response at the highest selected proton energy channel (15 MeV), while 9\% of them showed an irregular response.  2.5\% of shocks in the sample have a spike response in the 15 MeV channel, and none show a plateau-like response. For 72\% of the shock we have no associated energetic protons (both Zero flux and no response cases) in the 1 MeV channel, 13\% of them have an irregular response, 9\% have a spike and 6\% have a plateau response. 130 keV particles fluxes enhancements are found more easily at the observed shocks, with only 41\% of them not having a response. In this channel, 30\% of the shocks have irregular responses, 18\% of them have a spike response and the 11\% is associated with a plateau response. At the lowest energies studies with EPT (70 keV), 30\% of shocks were not associated with energetic protons intensity enhancements, and 31\%, 24\% and 14\% are associated with irregular, spike and plateau responses, respectively. Finally, we report on the responses for suprathermal ions with energies of 50 keV. For the 36\% of sample, we did not see any enhancement in the suprathermal population, while 33\%, 24\% and 6\% have irregular, spike and plateau responses, respectively. These results are compatible with the ones reported in \citet{Lario2003}. It is important to note that it is extremely common to observe irregular responses in the production of energetic particles for all the selected channels, once again emphasising the importance of considering the variability of the medium where shocks propagate, which  can profoundly modify the mechanisms of  their production and transport. The cross-correlation of the characterized profiles with shock parameters will be the object of future studies.

It is possible to extract further information about energetic particle production from the intensity--time profiles, by studying when particle fluxes peak relative to the shock crossing time~\citep[e.g.,][]{vanNes1984,Lario2003}. From Diffusive Shock Acceleration (DSA) theory, the time profiles are expected to peak at the same time as the shock crossing, a prediction confirmed by some observations, and failing incompatible with others~\citep[see][]{Giacalone2012, Perri2015, Kartavykh2016}.

It is then interesting, as a further use case of the database, to report on the time difference between the peak of energetic particles and the shock passage time. This characterization is provided in the SERPENTINE shock list for the same energy channels discussed above. Figure~\ref{fig:TmaxvsT} shows an overview of the time difference between the peak of energetic particle profiles and the shock crossing time for all 5 channels (left to right). Negative, zero and positive values of $\rm T_{Peak} - T_{Shock}$ are the cases when the peak in particle intensity occurs before, at and after the shock passage. In the Figure, the highlighted red area corresponds to 10 minutes centred around $\rm T_{Peak} - T_{Shock}= 0$, and therefore the points falling in the highlighted area have a response compatible with particle fluxes peaking at the shock (though small-scale departures are possible). Once again, Figure~\ref{fig:TmaxvsT} shows how complex cases where energetic particles peak in advance/with delay to the shock crossing time. This behaviour is due to several factors, including evolutionary effects of the shocks (i.e., shock parameters making acceleration more or less favoured are not constant in time), as well as spatial irregularities, where the spacecraft yields one-dimensional information sampling a complex, intrinsically three-dimensional environment. Finally, the fact that particle fluxes peaking later than the shock being more common than particles peaking in advance of the shock, suggests that particle trapping in downstream structures is important for further acceleration and for influencing particle transport properties~\citep[][]{Schwadron2020,Trotta2022a, Kilpua2023}, whereas intervening structures upstream may change this behavior~\citep[e.g.,][]{Lario2003}.

\section{Conclusions} \label{sec:conclusions}

This paper has presented the an extensive in-situ observational effort about IP shocks using Solar Orbiter. The importance of exploiting the Solar Orbiter dataset is twofold: on one hand, poorly explored heliocentric distances can be accessed; on the other hand, the unprecedented time-energy resolutions provided by the EPD suite open a new observational window in the study of energetic particles in the heliosphere, as shown by many recent efforts~\citep[e.g.,][]{Wimmer2021,Kollhoff2021,Trotta2023c,RodriguezGarcia2023}. This effort is in continuity with other studies using other missions and looking at interplanetary shocks statistics at different heliocentric distances~\cite [e.g.][]{Kilpua2015_shocks, Oliveira2023,PerezAlanis2023}.

A sample of 100 shocks was identified using the TRUFLS detection software (Appendix~\ref{appendix:TRUFLS}), publicly available and easily portable to other missions. All the shocks have been extensively characterized. The fundamental shock parameters were estimated using the publicly available SerPyShock code~\citep{Trotta2022b}. The outcome of the analysis, including also advanced information on the shocks such as the study of wave foreshocks and energetic particle response, is the object of the SERPENTINE shock list, free to download and use through a dedicated server developed in the framework of the project (Appendix~\ref{appendix:serpentine_list}). The SERPENTINE list employs a data processing pipeline where each shock is analysed with the same choice of free parameters in the diagnostic procedure. In particular, the length of the upstream/downstream averaging windows has been fixed between 1 and 8 minutes. In this work, we also give the shocks' basic parameters using a case-by-case procedure based on the visual inspection of each event, reported in Appendix~\ref{appendix:dimmock_list}.

The general trends of Solar Orbiter IP shocks have been presented and found to be compatible with previous studies~\citep[e.g.][]{Kilpua2015_shocks}. We found that IP shocks tend to be weak, most of them have only moderate Mach numbers ($\lesssim 3$). The analysis of shock normal vector distributions reveals that significant departures from radial are common, an important ingredient often overlooked when modelling shock propagation in the heliosphere. The shocks' gas and magnetic compression ratios, Alfv\'enic and fast magnetosonic Mach numbers, normal angles $\theta_{Bn}$ and speed have been studied as a function of heliocentric distance, revealing no strong correlation with this parameter. 
Comparing the shock parameters measured below 0.8 AU with the large sample of about 600 shocks at 1 AU provided in \citet{Kilpua2015_shocks}, we found that high Mach numbers are more common in the inner heliosphere. This can be due to an evolution effect of the shock drivers: at short heliocentric distances, most of the shocks are driven by CMEs, and they are faster in the early stages of their propagation~\citep{Vrsnak20210}. At short heliocentric distances, it is also easier to find quasi-parallel shocks compared to the 1 AU sample, an effect due to the Parker Spiral of the heliospheric magnetic field, being less curved at low heliocentric distances (an effect discussed in ~\citet{Chao1985,Reames1999}). Further studies on inner heliospheric shocks will involve both future Solar Orbiter shock observations and the integration of the present survey with a similar effort using the PSP mission (see the SODA database \url{https://parker.gsfc.nasa.gov/shocks.html}).

About 50\% of the shocks show enhanced wave activity, as revealed by magnetic field wavelet analyses performed on the entire shock sample. The identified wave foreshocks in frequencies between 0.01 and 1 Hz last from a few minutes to about one hour. Precursors at higher frequencies follow similar statistics. The analysis of the shocks' wave environment indicates an emerging complexity in the interplay between shock-generated and pre-existing magnetic fluctuations. The wave foreshocks identified are often associated with shock-reflected particles, less easily identified due to the plasma instrument's SWA-PAS instrumental limitations. Long-lasting (about 1 hour) wave/particle foreshocks have been found upstream of shocks propagating in the solar wind with low levels of magnetic field fluctuations, a result compatible with recent studies of long-lasting field-aligned beams of particles with higher energies~\citep{Lario2022}. Interestingly, for some shocks, enhanced wave activity has been identified without particle counterparts. These, firstly reported by \citet{Kajdic2012}, will be the object of future studies.

In this work, we also gave an overview of the novel capabilities of the Solar Orbiter EPD suite, yielding detailed information about energetic particle behaviour at shocks and often showing complexity in particle production, beyond simple acceleration models~\citep{Yang2024, Trotta2023c}. Energetic protons' time-intensity profiles and their response to the passage of IP shocks were characterised for 5 different energy channels,  (15 MeV, 1 MeV, 130 keV, 70 keV and 50 keV). About 70\% of the shocks were found to be associated with 50 keV proton intensity increases. In contrast, 15 MeV particles were associated with 10\% of the shocks showing all irregular response, i.e. there were no plateau cases  at these high energies that are considered as DSA-like response. We note that many shocks have irregular/complex particle responses for all the energy channels, highlighting how the ambient fluctuations and shock irregularities shape particle production and transport features. Peak times of energetic particles have also been studied relative to the shock crossing times, revealing that particles may often peak earlier/later than the time the shock crosses the spacecraft. This is another effect beyond the classical picture of particle acceleration at shocks. This study is complementary to previous surveys of shock accelerated particles~\citep{Lario2003} and, in future studies, will be put in the context of the variability found in particle time-intensity profiles using a multi-mission approach~\citep{Neugebauer2006}. In future work, the correspondence between energetic particles' response and resonant wavelength magnetic field fluctuations, a fundamental ingredient of shock acceleration theories, will be investigated, building onto earlier work done with other missions and for a narrower range of energies~\citep{Desai2012}.

\section{Acknowledgments}
This study has received funding from the European Union's Horizon 2020 research and innovation program under grant agreement No. 101004159 (SERPENTINE, www.serpentine-h2020.eu). Views and opinions expressed are, however, those of the authors only and do not necessarily reflect those of the European Union or the European Research Council Executive Agency. Neither the European Union nor the granting authority can be held responsible for them. This work was supported by the UK Science and Technology Facilities Council (STFC) grant ST/W001071/1.
 Solar Orbiter magnetometer operations are funded by the UK Space Agency (grant ST/X002098/1). Solar Orbiter is a space mission of international collaboration between ESA and NASA, operated by ESA. Solar Orbiter Solar Wind Analyser (SWA) data are derived from scientific sensors which have been designed and created, and are operated under funding provided in numerous contracts from the UK Space Agency (UKSA), the UK Science and Technology Facilities Council (STFC), the Agenzia Spaziale Italiana (ASI), the Centre National d'Etudes Spatiales (CNES, France), the Centre National de la Recherche Scientifique (CNRS, France), the Czech contribution to the ESA PRODEX program and NASA. Solar Orbiter SWA work at UCL/MSSL is currently funded under STFC grants ST/W001004/1 and ST/X/002152/1. The Energetic Particle Detector (EPD) on Solar Orbiter is supported by the Spanish Ministerio de Ciencia, Innovación y Universidades FEDER/MCIU/AEI Projects ESP2017-88436-R and PID2019-104863RB-I00/AEI/10.13039/501100011033 and the German space agency (DLR) under grant 50OT2002. X.B.-C. is supported by DGAPA-PAPIIT grant IN106724. H.H. is supported by the Royal Society University Research Fellowship URF R1 180671. N.D. is grateful for support by the Academy of Finland (SHOCKSEE, grant No. 346902). T.S.H. is supported by STFC grant ST/W001071/1. 

%



\clearpage
\appendix
\section{The TRUFLS shock identification algorithm}\label{appendix:TRUFLS}

The Tracking and Recognition of Universally Formed Large-scale Shocks (TRUFLS) code is built to look at long time
series and identify shocks automatically. An important, similar effort has been done for other missions in the
Heliospheric Shock Database generated and maintained at the University of Helsinki; see \url{http://ipshocks.fi} for further details. To compile such catalogue, the authors used either visual inspection 
of magnetic field and plasma data, or, for a small number of missions, a machine learning algorithm 
(InterPlanetary Support Vector Machine, IPSV, \url{https://pypi.org/project/ipsvm/}). Once a shock candidate was 
identified through one of the two methods, the authors required a set of upstream/downstream relations on the 
magnetic field plasma data to be satisfied in order to confirm that the candidate was indeed a shock. 

TRUFLS, instead, looks where such jump conditions are satisfied using a moving average scanning the entire 
timeseries that is required to analyse. The conditions to be satisfied to identify a shock event are the 
following:
\begin{eqnarray}
     \mathrm{\frac{B_d}{B_u} \geq 1.2} \label{eq:cond1} \\
     \mathrm{\frac{n_d}{n_u}\geq 1.2} \label{eq:cond2} \\
     \mathrm{FF: \,\,\,\, V_d - V_u  \geq 20 \,\, km/s } \label{eq:cond3} \\
     \mathrm{FR: \,\,\,\, V_u - V_d  \geq 20 \,\, km/s } \label{eq:cond4}
\end{eqnarray}
Here, the subscripts $\mathrm{u}$ and $\mathrm{d}$ indicate upstream and downstream averages respectively, and 
$\rm B$ and $\rm V$ denote the magnetic field magnitude and spacecraft frame plasma flow speed, respectively. 
Equations~(\ref{eq:cond1}) and~(\ref{eq:cond2}) represent the compression of magnetic field and plasma density 
expected at the shock, respectively. The criterion on the plasma bulk flow speed for the Fast Forward (FF) and 
Fast Reverse (FR) cases are summarised in Equations~(\ref{eq:cond3}) and (\ref{eq:cond4}), respectively. We 
restrict our analyses on fast shocks (i.e., shocks for which the shock speed is larger than the upstream fast magnetosonic speed). Vice versa, slow shocks are not treated here. It is worth underlining that in the ipshocks.fi database, a further constraint is requested to confirm that the candidate event is indeed a shock, namely the proton temperature jump $\mathrm{T_d/T_u} \geq 1.2$. Within our identification, we relax this request, due to the fact that temperature data are the ones with the highest levels of noise.  
\begin{figure}[]
    \centering
    \includegraphics[width=\textwidth]{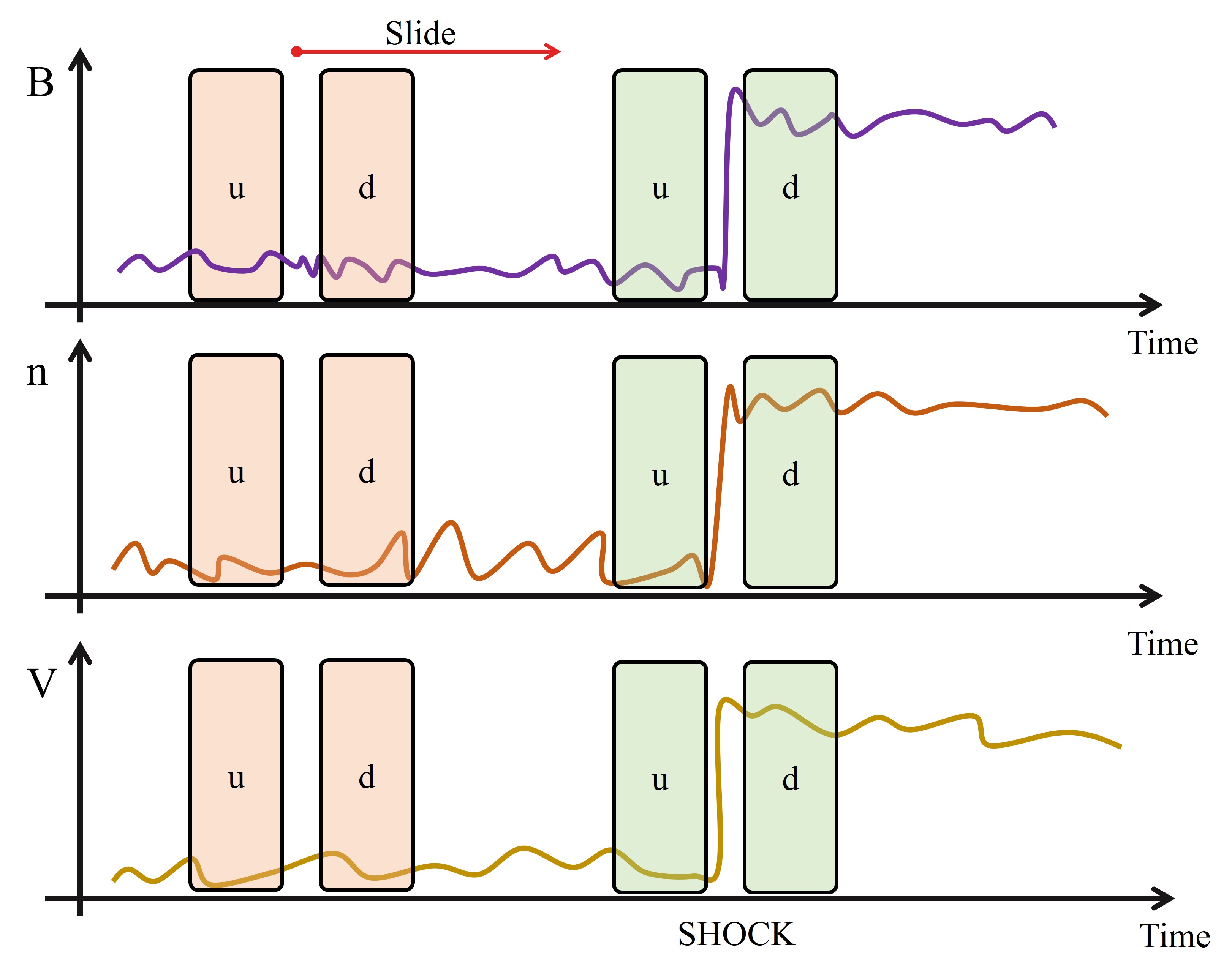}
    \caption{Sketch showing the sliding average windows (u and d, shaded panels) identifying a FF shock (green) as done in the TRUFLS code.}
    \label{fig:trufls_sketch}
\end{figure}

The TRUFLS code searches for times where Equations~(\ref{eq:cond1}--\ref{eq:cond4}) are verified in time series 
containing magnetic field and plasma data. Upstream and downstream averaging windows (and an exclusion zone where 
the shock itself will be) are chosen, with their length being a user-defined parameter. The output of the code 
consists of the times when the jump conditions are satisfied, that therefore constitute the shock candidates. The 
process by which TRUFLS operates is further elucidated in Figure~\ref{fig:trufls_sketch}, where a sketch of a FF 
shock signature is reproduced together with the sliding averaging windows. 

It is worth underlining that when plasma data is not available, such diagnostics become more complex. While we 
developed a TRUFLS version working on magnetic field only candidates, that flags a lot of false positives, as 
Equation~(\ref{eq:cond1}) is easily satisfied also at structures that are not shocks. For this reason, to 
automatically identify shocks, we strongly recommend using both plasma and magnetic field data.

The TRUFLS distribution is publicly available at \url{https://github.com/trottadom/PyTRUFLS}, and it may be used to identify new Solar Orbiter shocks with the most recent dataset, as well as on datasets from other missions.

\section{The Solar Orbiter shock list in the SERPENTINE data centre}
\label{appendix:serpentine_list}

The Solar Orbiter shock catalog presented here has been produced and developed as part of the  Solar Energetic particle analysis platform for the inner heliosphere (SERPENTINE) project. SERPENTINE targeted three main scientific questions, namely:
\begin{itemize}
    \item Q1: What are the primary causes for widespread SEP events observed in the heliosphere? \\
    \item  Q2: What are the shock acceleration mechanisms responsible for accelerating ions from thermal/suprathermal energies to near-relativistic energies in the corona and in the interplanetary medium? \\
    \item Q3: What is the role of shocks in electron acceleration in large gradual and widespread events? How does it relate to ion acceleration and what is its importance relative to flare acceleration?
\end{itemize}

Scientific closure of Q1-Q3 has been targeted through comprehensive analyses, both case studies and statistical investigations, of historical and current SEP measurements and solar context observations \citep{Kollhoff2021, Dresing2022, Dresing2023, Rodriguez-Garcia2023a, Rodriguez-Garcia2023b, Afanasiev2023, Wijsen23,  Jebaraj2023, Jebaraj2023b, Trotta2022a, Dimmock2023, Lorfing2023, TalebpourSheshvan2023, Kilpua2023, Pezzi2023, Trotta2023b, Trotta2023c, Trotta2024, Trotta2024b, Hietala2024, Jebaraj2024, Wei2024, Khoo2024, Morosan2024}. 

\begin{figure}[]
    \centering
    \includegraphics[width=\textwidth]{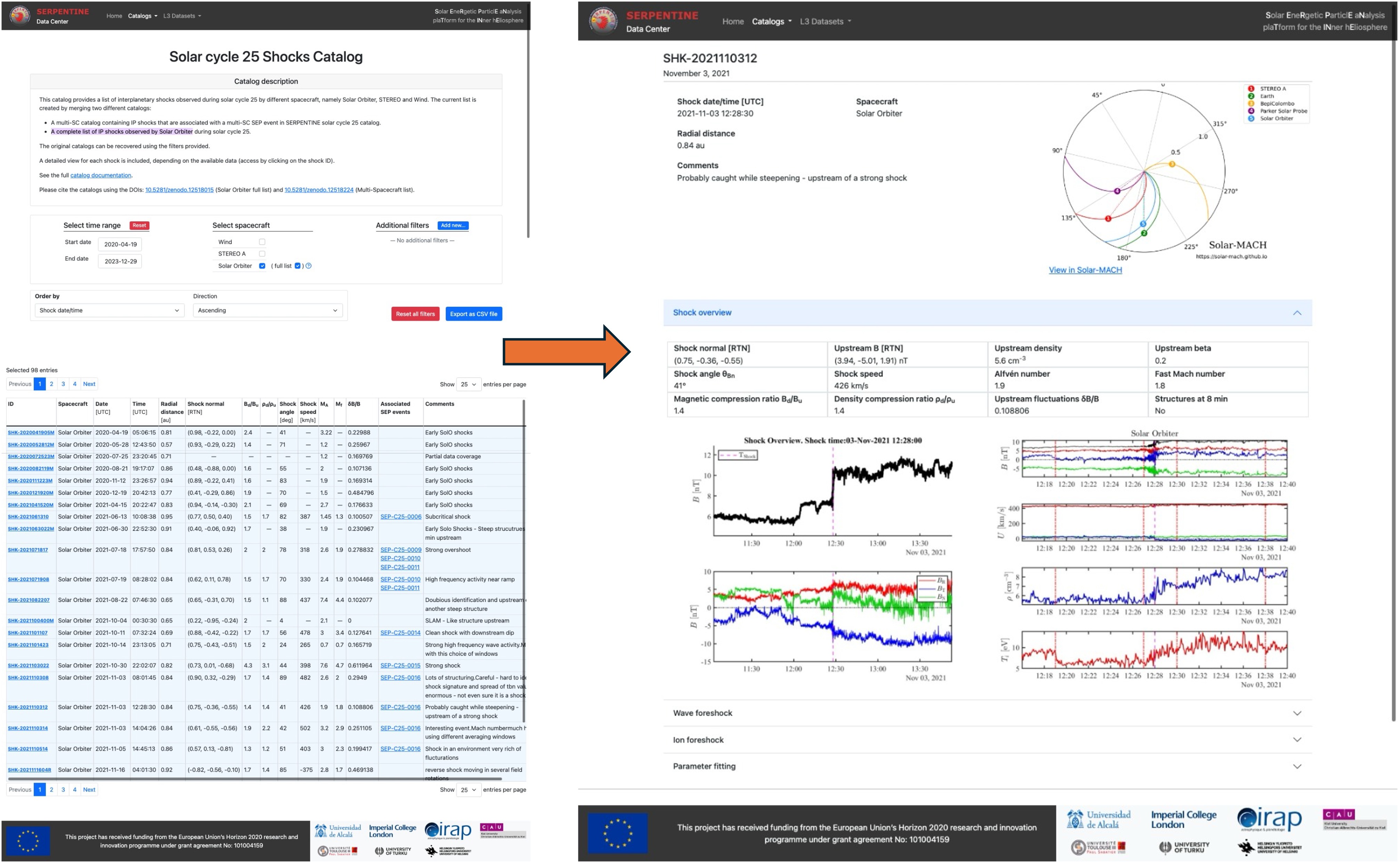}
    \caption{Screenshots of the Cycle 25 Solar Orbiter shock list in the SERPENTINE data centre (left), with advanced information and quicklook plots on one example event (right). Available at \url{https://data.serpentine-h2020.eu/catalogs/sep-sc25/}. }
    \label{fig:list_screenshot}
\end{figure}

Crucially, SERPENTINE has produced and delivered a very large public release of diverse data analysis tools, \citep{Kouloumvakos2022b, Palmroos2022, Price2022, Trotta2022b, Gieseler2023, Kouloumvakos2023} and catalogs of SEP events, IP shocks, and CMEs for historical and solar cycle 25 events. The tools and catalogs were built for easy access and are released in the hope of broad use by the heliophysics community. Altogether, five catalogs, two based on Helios data and three based on modern observations, have been released through the project data server.\footnote{\url{https://data.serpentine-h2020.eu}}. Here, we focused on the in-situ Cycle 25 shocks of Solar Orbiter, citable in the present version \citet{Trotta2024CatalogueZenodo} through Zenodo.\footnote{\url{https://zenodo.org/records/12518015}}. The list contains all the informations listed in Table~\ref{tab:table_list} as well as quicklook plots and links to the SEP catalogue of SERPENTINE~\citep{Dresing2024list}. An example of how to access and use the shock list is shown in Figure~\ref{fig:list_screenshot}, where the main page displaying events in the catalog is shown on the left. By clicking on each event, it is possible to access advanced information (spacecraft configuration, quicklook plots etc.), as shown by the panel on the right. Further information, like quicklook plots of the wave environment and energetic particle response may be accessed by clicking on the relevant sections.

\section{Case-by-case shock parameters estimation through visual inspection}
\label{appendix:dimmock_list}

The Solar Orbiter shock list was also inspected case-by-case with a semi-automated approach is adopted. Firstly, the IP shocks are identified automatically, using the same approach as the Heliospheric Shock Database generated and maintained at the University of Helsinki; see \url{http://ipshocks.fi}. This provides a list of IP shock candidates, and the following procedure is applied.
\begin{itemize}
    \item Inspect each candidate and remove events where the field and plasma parameter changes are inconsistent with fast mode interplanetary shocks (fast or reverse). Most misidentifications are current sheets or other complex structures.
    \item Merge windows when the same shock is identified multiple times to create a larger window that extends from upstream to downstream.
    \item For each candidate, the upstream and downstream regions are inspected to find the most reliable windows to compute shock parameters. We select the regions closest to the shock front that do not contain significant field rotations or changes in plasma moments.
    \item Shock parameters are computed using mean averages of the upstream windows. The shock normal uses the mixed mode coplanarity method, and the shock speed is computed according to mass flux conservation. 
    \item Parameters are checked to see if they are reasonable considering the structure of the shock. For example, a shock with $\theta_{bn} \sim 90^\circ$ would not be expected to exhibit a foreshock.
\end{itemize}

It is worth noting that some additional shocks existed in this database when the automatic identification was not possible, for example, early in the mission when no ion moments were available due to the operation of the SWA-PAS instrument. These shocks were identified by looking for clear shock signatures (ramp, foot, overshoot, compression) in the MAG data and electron density calibrated from the spacecraft potential. This final database is freely available \cite{dimmock_2024_14001054}.


\clearpage

\bibliography{bibby}{}
\bibliographystyle{aasjournal}



\end{document}